\documentstyle[multicol,aps]{revtex}
\begin{document}
\draft
\title{Fractional Exclusion Statistics and the Universal
Quantum of Thermal Conductance: A Unifying Approach}
\author{Luis G. C. Rego and George Kirczenow}
\address{Department of Physics, Simon Fraser University,
Burnaby, B.C., Canada  V5A 1S6}
\maketitle
\begin{abstract}

We introduce a generalized approach to one-dimensional (1D)
conduction based on Haldane's concept of fractional
exclusion statistics (FES) and the Landauer formulation of
transport  theory. We show that the 1D ballistic thermal
conductance  is independent of the statistics obeyed
by the carriers and is governed by the universal quantum
$\kappa^{univ}=\frac{\pi^2}{3}\frac{k_B^2 T}{h}$ in the
degenerate regime.  By contrast, the electrical
conductance of FES systems is statistics-dependent. This
work unifies previous theories of electron and phonon
systems and explains an interesting commonality in their
behavior.

\end{abstract}

\begin{multicols}{2}

\section{Introduction}

Recent theoretical investigations of quantum transport have
revealed an intriguing commonality in the behavior of some
apparently very dissimilar systems: It has been predicted that in
one dimension the low temperature ballistic thermal conductances
of ideal electron gases,\cite{Gut,extra} of phonons,\cite{Reg}  and of
interacting electrons that form chiral\cite{Kan} or
normal\cite{Fazio} Luttinger  liquids should all
be quantized in integer multiples of a universal quantum
$\kappa^{univ}=\frac{\pi^2}{3}
\frac{k_B^2 T}{h}$, where $T$ is the temperature, $k_B$ the
Boltzmann constant and $h$ is Planck's constant. That is, a 1D
band populated with bosons described by a Planck distribution
(phonon modes) has been predicted to transport the {\em same}
amount of heat as one populated by fermions (the ideal electron
gas) or a Luttinger liquid. Also, experimental evidence has been
reported that ropes of single walled nanotubes conduct heat in
amounts proportional to $\kappa^{univ}$ \cite{Hone}. However each
of these systems was studied separately using a different
theoretical approach. Thus it has been unclear whether this
convergence of the results that have been obtained is simply a
coincidence or whether it has a deeper significance and broad
ramifications. The purpose of this work is to resolve this
question with the help of the concept of fractional exclusion
statistics (FES), proposed by Haldane\cite{Hal}, that allows one
to discuss the behavior of bosons, fermions and particles having
fractional statistical properties, all on the same footing.
Besides the universal thermal conductance, we also obtain
naturally from this theory the quantized  electrical conductance
for ballistic electrons in 1D quantum wires  and in the fractional
quantum Hall (FQH) regime.

FES\cite{Hal} extends the concept of anyons,\cite{Wil} i.e.,
particles with fractional statistics, from two dimensions to
arbitrary spatial dimensions by introducing a generalization of
the Pauli exclusion principle, and has yielded novel insights into
fractional quantum Hall systems,\cite{Hal,Elb} spinons in
antiferromagnetic spin chains\cite{Hal}, systems of interacting
electrons in 2D quantum dots,\cite{Bha} and the
Calogero-Sutherland model.\cite{Mur1}  In Haldane's sense the
statistics of a system composed of different species of  particles
(or quasi-particles) is defined by the relation
$\Delta d_i=-\sum_j g_{ij} \Delta N_j$, where $N_i$ is the number
of particles of species $i$ and $d_i$ is the dimension of the
$N_i$-particle Hilbert space, holding the coordinates of the
$N_i-1$ particles fixed.  The parameter $g_{ij}$ is the
statistical interaction. For a system of identical particles {\it
g} is a scalar quantity, with $g=1$ (0) for fermions  (bosons). Wu
\cite{Wu} has used this definition of FES to establish the
statistical distribution function for an ideal gas of particles
with fractional statistics. It has been proposed that such ideal
FES gases provide an accurate representation of the physics of a
number of interacting electron systems.\cite{Bha,Mur1}

While much attention has been given to the thermodynamic
properties of FES systems, \cite{Mur1,Wu,Nay,Raj,Huang1,Igu} their
transport properties have not received the same consideration. In
this paper we use the Landauer formulation of transport
theory\cite{Lan} to study  conduction in ideal one-dimensional FES
systems. Remarkably, we find that their low temperature thermal
conductance is quantized in integer multiples of the universal
quantum $\kappa^{univ}=\frac{\pi^2}{3}
\frac{k_B^2 T}{h}$, {\em irrespective of the value of the
statistical parameter} $g_{ij}$. Thus we demonstrate that the
quantization of thermal conductance and the associated quantum are
statistics-independent and truly universal.
By contrast we find the electrical conductances of FES systems to be 
statistics-dependent.

\section{Single Species}
\label{ss}

Consider a two terminal transport experiment where two  infinite
reservoirs are adiabatically connected to each other by an
one-dimensional channel. Each reservoir is characterized by a
temperature ($T$) and a chemical potential ($\mu$), considered to
be independent variables. In the case of reservoirs with charged
particles $\mu$ can be redefined as the electrochemical potential,
that is, a combination of the chemical potential and an
electrostatic particle energy governed by an external field. In
terms of
$T$ and $\mu$ the electric ($I$) and energy ($\dot{U}$)  currents
in the linear response regime are
\begin{eqnarray}
\delta I &=& \left. \frac{\partial I}{\partial \mu} \right|_T \delta\mu
          +  \left. \frac{\partial I}{\partial T} \right|_{\mu} \delta T
\label{e1}
\\
\delta \dot{U} &=& \left. \frac{\partial \dot{U}}{\partial \mu} \right|_T
                   \delta \mu
                +  \left. \frac{\partial \dot{U}}{\partial T} \right|_{\mu}
                   \delta T \ ,
\label{e2}
\end{eqnarray}
where $\delta T = T_R - T_L$ and $\delta \mu = \mu_R - \mu_L$, with
$R$ ($L$) representing the right (left) reservoir.

Using Landauer theory we write the fluxes between the
two reservoirs as
\begin{eqnarray}
I &=& \sum_{n} \ q \int_0^{\infty} \frac{dk}{2\pi}  \ v_n(k)
     \ \left[\eta_R -\eta_L \right]\ \zeta_n(k)
\label{e3}
\\
\dot{U} &=& \sum_{n} \int_0^{\infty} \frac{dk}{2\pi} \varepsilon_n(k)
           \ v_n(k) \left[ \eta_R - \eta_L \right] \ \zeta_n(k) \ .
\label{e4}
\end{eqnarray}

The sum over $n$ takes into account the independent propagating modes
admitted by the channel. $\varepsilon_{n}(k)$ and
$v_{n}(k)$ are the energy and velocity of the particle with wave-vector
$k$, $\zeta_{n}(k)$ is the particle transmission probability through the
channel, $\eta_i$ represents the statistical distribution functions
in the reservoirs and $q$ is the particle charge.
In one-dimension  the particle velocity
$v_n(k) = \hbar^{-1}(\partial \varepsilon_n/\partial k)$ is canceled by the
1D density of states
${\cal D} (\varepsilon_n)=\partial k/\partial\varepsilon_n$
and the fluxes become independent of the dispersion
\begin{eqnarray}
I &=& \frac{q}{h} \sum_{n} \ \int_{\varepsilon_{n}(0)}^{\infty}
      d\varepsilon \ \left[ \eta_R - \eta_L \right] \zeta_n(\varepsilon)
\label{e5}
\\
\dot{U} &=& \frac{1}{h} \sum_{n} \ \int_{\varepsilon_{n}(0)}^{\infty}
          d\varepsilon\ \varepsilon \left[ \eta_R- \eta_L \right]
          \zeta_n(\varepsilon) \ .
\label{e6}
\end{eqnarray}
Throughout the remainder of this
paper we will assume $\zeta_n(\varepsilon)=1$,
which
corresponds to ballistic transport and a perfectly
adiabatic coupling between the
reservoirs and the 1D system. This assumption
can be considered realistic
in view of the present stage of the mesoscopics technology.

Substitution of expressions (\ref{e5}) and (\ref{e6})
for the fluxes into Eqs.(\ref{e1}) and
(\ref{e2}), while taking the limit $\delta T \rightarrow 0$ and $\delta
\mu \rightarrow 0$, gives us the transport coefficients.

Having introduced the model, we consider
systems of generalized statistics,
which can be investigated within
FES theory.
Initially we concentrate on identical particle systems
and the distribution function derived by Wu \cite{Wu} for an ideal gas
of particles obeying FES
\begin{eqnarray}
\eta_g = \frac{1}{{\cal W}(x,g)+g}
\label{e12}
\end{eqnarray}
with $x \equiv \beta(\varepsilon -\mu)$, $\beta \equiv 1/(k_B T)$ and
${\cal W}(x,g)$ given by the
implicit equation
\begin{eqnarray}
{\cal W}^g(x,g)[1+{\cal W}(x,g)]^{1-g} = e^{x} \ .
\label{e13}
\end{eqnarray}
Making $g=0$ or $g=1$ Eq.(\ref{e12}) becomes the Bose-Einstein or
the Fermi-Dirac distribution function, respectively.

For a system of generalized statistics the transport coefficients are
\begin{eqnarray}
L_{11} &=& \left. \frac{\partial I}{\partial \mu} \right|_T
        = \frac{q}{h} \sum_{n} \int_{x_{0n}}^{\infty} dx\ F(x,g)
\label{e14}
\\
L_{12} &=& \left. \frac{\partial I}{\partial T} \right|_{\mu}
        =  \frac{q}{h}k_B \sum_{n} \int_{x_{0n}}^{\infty} dx\ x\ F(x,g)
\label{e15}
\\
L_{21} &=& \left. \frac{\partial \dot{U}}{\partial \mu} \right|_T
        = \frac{1}{h\beta} \sum_{n} \int_{x_{0n}}^{\infty}
         dx\ (x + \mu\beta) \ F(x,g)
\label{e16}
\\
L_{22} &=& \left. \frac{\partial \dot{U}}{\partial T} \right|_{\mu}
        =  \frac{k_B}{h\beta} \sum_{n} \int_{x_{0n}}^{\infty}
           dx\ (x^2+x\mu\beta)\ F(x,g)
\label{e17}
\end{eqnarray}
with $x_{0n} \equiv \beta(\varepsilon_{n}(0) -\mu)$ and
\begin{eqnarray}
F(x,g) = \frac{{\cal W}(x,g)[{\cal W}(x,g) + 1]}{[{\cal W}(x,g) + g]^3} \ .
\label{e18}
\end{eqnarray}

Fermions and bosons are the special cases of the theory,
however, our interest
is to develop a formalism able to treat all FES systems.
Analytic solutions of Eq.(\ref{e13}) can also be obtained for the
special cases $g=\frac{1}{4},\frac{1}{3},\frac{1}{2},2,3$ and 4, but for
general $g$ the approach of  analytically solving the equation
for ${\cal W}$ is not
possible. We now present a comprehensive method to treat
this problem. Initially, solve Eq.(\ref{e13}) for $x=\beta(\varepsilon-\mu)$
\begin{eqnarray}
x({\cal W},g) = ln({\cal W}+1) + \left[ ln({\cal W}) - ln({\cal W}+1) \right]
g \ .
\label{ex}
\end{eqnarray}
We notice that $\lim_{{\cal W}\rightarrow 0} x = - \infty$
for $g \not = 0$,
which corresponds to the lowest energy for the degenerate non-bosons.
Moreover, $\lim_{{\cal W}\rightarrow 0} x = 0$ when $g=0$,
that corresponds to the lowest energy modes of bosons
described by the Planck distribution.
On the other hand, $\lim_{{\cal W}\rightarrow \infty} x = \infty$
for any $g \geq 0$.
This shows that $x$ can be supplanted by $\cal{W}$ as the variable
of integration in our general FES expressions
for $L_{ij}$, with $\cal{W}$ ranging from
0 to infinity.
Notice that no other specification on the functional form of the
particle spectra is made in this derivation.
Then, using Eq.(\ref{ex}), we can write
\begin{eqnarray}
F(x,g)dx = \frac{d{\cal W}}{({\cal W} + g)^2} \ .
\label{FW}
\end{eqnarray}
The transport coefficients $L_{ij}$ can then be evaluated analytically
for arbitrary $g$.  When $g > 0$
\begin{eqnarray}
L_{11} &=& M \frac{q}{h} \ \int_0^{\infty} \frac{d{\cal W}}{({\cal W} + g)^2}
        =  M \frac{q}{h}\ \frac{1}{g}
\label{e34}
\\
L_{12} &=& M \frac{q}{h}k_B \int_0^{\infty} d{\cal W}
           \ \frac{x({\cal W},g)}{({\cal W} + g)^2} = 0
\label{e35}
\\
L_{21} &=& M \frac{1}{h\beta}\ \int_0^{\infty} d{\cal W}
           \ \frac{x({\cal W},g) + \mu\beta}{({\cal W} + g)^2}
        =  M \frac{\mu}{h}\ \frac{1}{g}
\label{e36}
\end{eqnarray}
and for all $g \ge 0$
\begin{eqnarray}
L_{22} &=& M \frac{k^2_B T}{h}\ \int_0^{\infty} d{\cal W}
           \ \frac{x^2({\cal W},g)+\mu\beta x({\cal W},g)}{({\cal W} + g)^2}
\nonumber \\
       &=& M \frac{k^2_B T}{h}\ \frac{\pi^2}{3}
\label{e37}
\end{eqnarray}
with $x({\cal W},g)$ given by Eq.(\ref{ex}).
Here $M$ is an integer number that takes into account
the number of occupied modes (assuming a degenerate population in each
one of them).
Therefore the  transport equations for a system of identical  particles
of generalized statistics are
\begin{eqnarray}
\delta I &=& \frac{1}{g} \frac{q}{h} M \delta \mu
\label{e27}
\\
\delta \dot{U} &=& \frac{1}{g} \frac{\mu}{h} M \delta \mu +
\frac{\pi^2}{3}\frac{k_B^2 T}{h} M \delta T \ .
\label{e28}
\end{eqnarray}
One important result that
we obtain with this formalism is the universal thermal
conductance, {\it valid for all ballistic FES systems}.
Since the electro-chemical potential is an
independent variable in this model, we can
set $\delta \mu =0$ so that no electric current flows between the reservoirs.
This also eliminates the energy flow that is due to a net flux of
particles between the two reservoirs, leaving us
with only the coefficient $L_{22}$. In this case the energy current
is equal to the
heat current that is generated by $\delta T$,
and so the 1D universal thermal conductance is
\begin{eqnarray}
\kappa^{univ} = \frac{\pi^2}{3}\frac{k_B^2 T}{h}  \ .
\label{univ}
\end{eqnarray}
Therefore a 1D subband populated with bosons described by the Planck
distribution transports the same amount of heat as one populated by
fermions, despite the fact these systems have
very different statistical behaviors. 
The thermopower vanishes because of the assumptions made:
degenerate systems and unitary transmission coefficients
independent of the
energy. For Planck bosons
$\mu$ is not a parameter describing the system, therefore only $L_{22}$ is
present and  the result (\ref{univ}) is recovered.

We note that, in contrast to the thermal conductance, the 1D ballistic
electrical conductance is {\it not} statistics-independent since $g$ appears
explicitly in Eq.(\ref{e27}) for the electric current.
For instance when $\delta T = 0$, the fermion case is readily obtained
by setting $g=1$ and we obtain the well known 1D electrical conductance
$G = (e^2/h)M$ for ballistic electrons.

Equations (\ref{e27}) and (\ref{e28}) should also
describe the transport properties
of the Laughlin states of the FQHE, for which
the Landau level filling fraction
$\nu = 1/(2m+1)$ with m integer. We use the composite fermion (CF) picture
\cite{Jain1} to derive the statistical interaction parameter {\it g}
of these particles. Integrating
the expression $\Delta d = - g \Delta N$ for
$g = g^{CF}$ we get $d_N = d_0^{CF} - g^{CF}(N-1)$, where
$d_N$ is the dimension of the one particle Hilbert space when $N$ composite
fermions exist in the system whereas $d_0^{CF}$ is its analog in the absence
of CF. The term $d_0^{CF}$ is the degeneracy of the CF
Landau level, that can be
written in terms of the CF density as $d_0^{CF} = (eB/hc) - 2mN$,
with $B$ representing the external magnetic field. Using this relation
in the expression for $d_N$ along with the fact that the CF behave like
fermions ($g^{CF} = 1$) we get $d_N = (eB/hc) - (2m+1)N + 1$. This
means that, from the perspective of the FES theory,
the transport properties of these states are due to  particles of
charge $|q|=e$ and a fractional exclusion rule given by $g=(2m+1)$
in the thermodynamic limit.
In other words, each electron added to the system excludes $2m+1$
single particle states.
Returning to the transport equations, substitution of $g=(2m+1)$ in
Eq.(\ref{e27}) gives us immediately the well known values of the conductance
plateaus of the Laughlin states, $G = \frac{1}{2m+1}\frac{e^2}{h}$.
Since on the FQH plateaus the two-terminal conductance is equal to the
quantized Hall conductance, this result is in agreement with experimental
data \cite{Ts} on FQH devices.
It is important to mention that, since we are concerned with transport,
our analysis applies to the electrons themselves as distinct
from the quasi-particle excitations studied in reference \cite{Wu98},
that was concerned with thermodynamics.

The universality presented by the degenerate 1D systems at
finite $T$ can be physically understood if we consider the total energy
flux for a single band
\begin{eqnarray}
\dot{U} = \frac{\mu^2}{2gh} + \frac{\pi^2}{6}\frac{(k_B T)^2}{h}
        = \dot{U}_{pot} + \dot{U}_{thermal}\ ,
\label{U}
\end{eqnarray}
which shows that the energy current flowing
through the one-dimensional system
can be divided in two independent components:
one due exclusively to the flux
of particles and carrying no heat ($\dot{U}_{pot}$)
and the other entirely determined
by the temperature of the emitting reservoir
irrespective of the number of particles ($\dot{U}_{thermal}$).
The last term gives rise to the thermal conductance being the same
for Planck bosons and all other FES particles.
This division is possible because of the cancellation of the density of
states by the particle velocity in the 1D system
along with the degenerate condition
of the system. On the other hand, the electric current for degenerate
systems depends only on the
number of particles regardless their temperature,
which leads to $L_{12}=0$.

\section{Generalized Exclusion and FQHE}

In the previous section we have shown that the generalized exclusion 
approach for a system of identical particles leads naturally to the 
transport coefficients of the Laughlin fractions of the FQHE.
In the remainder of this paper we extended this formalism to treat
systems composed 
of multiple species with a mutual statistical interaction acting among 
them.

\subsection{Exclusion Statistics for various species}

In its most general form the occupation numbers $\eta_i$ of each species
that assembles into an ideal gas of FES particles are given by
\begin{eqnarray}
{\cal W}_i = \frac{1}{\eta_i} - \sum_{j=1}^S g_{ij} \frac{\eta_j}{\eta_i}
\label{multi1}
\end{eqnarray}
and
\begin{eqnarray}
(1+{\cal W}_i) \prod_{j=1}^S \left( \frac{{\cal W}_j}{{\cal W}_j + 1}
\right)^{g_{ji}}
= e^{x_i} \ ,
\label{multi2}
\end{eqnarray}
where $x_i = \beta_i(\epsilon_i - \mu_i)$ and $S$ is the number of species. 
Details of this derivation can be found in reference \cite{Wu}.

To proceed with the construction of the statistics of the model it is 
convenient
to introduce the actual values of $g_{ij}$. To do so we use Jain's
Composite Fermion picture \cite{Jain1} and the generalized exclusion
principle
\begin{eqnarray}
G_{eff,i} = G_i - \sum_j g_{ij} (N_j - \delta_{ij}) \ .
\label{w5}
\end{eqnarray}

The main properties of the FQHE can be understood if
we attach an even number of fictitious flux quanta to each single electron by
a Chern-Simons  gauge transformation \cite{Lopez}. 
In this picture a dressed particle is formed
which has the same  charge and the statistical properties of the electron.
In the mean field approximation the CFs form a Fermi liquid.
From this perspective the FQHE is then seen as the IQHE of the CF particles,
which experience an {\it effective} magnetic field that depends on the
density of carriers $B_{eff} = B - 2mN\Phi_0$, where
$B$ is the external magnetic field, $\Phi_0 = ch/e$ is the quantum of
magnetic flux and $N$ is the total density of CFs
in the mean field approximation (the same as the electronic density).
Therefore, according to this picture, the quasi-Landau levels (qLLs) 
occupied by the CF have a degeneracy that is
\begin{eqnarray}
G^{CF} = B_{eff}/\Phi_0 = B/\Phi_0 - 2m\sum_{j=1}^S N_j 
\label{cfdeg}
\end{eqnarray}
with the index $j$ representing the qLL index.
Moreover, because CFs are fermions we have
$g_{ij}^{CF} =\delta_{ij}$ and expression (\ref{w5})
can be written as
\begin{eqnarray}
G_{eff,i} &=& G^{CF}_i - \sum_j g^{CF}_{i,j}(N_j-\delta_{ij})
\nonumber \\
&=& \frac{B}{\Phi_0} - \sum_j (2mN_j + \delta_{ij}) + 1 \ .
\label{2s5}
\end{eqnarray}
Regrouping the elements according to the densities $N_j$ of each
qLL we obtain $g_{ii} = 2m + 1$ and $g_{ij} = 2m$, for
$i \not = j$. In the FES theory the diagonal terms are the self-interaction
parameters and rule the exclusion properties among particles of the same
species whereas the non-diagonal terms are the statistical 
mutual interaction parameters
which describe the exclusion relations among particles of different species.
In this case the population of each qLL is viewed as
a distinct species.
What expression (\ref{2s5}) shows is that we can incorporate the physics
of Jain's fractions $\nu = \frac{p}{2mp+1}$ ($m$ and $p$ integers) into the
generalized exclusion principle of particles in the FES theory.
These new particles have the same charge as the electron ($|q|=e$), but their
exclusion statistics is governed by $g_{ij}$.

The knowledge of $g_{ij}$ allows us to solve equation (\ref{multi1}) to
obtain the occupation functions $\eta_i$ for each species. For a system
composed of a number $S$ of  species that obey the exclusion rules derived
above we have
\begin{eqnarray}
\eta_i = \frac{\prod_{j\not=i}^S ({\cal W}_j + 1)}
{\prod_{j=1}^S ({\cal W}_j + 2m + 1) - \Lambda} \ ,
\label{etai}
\end{eqnarray}
with
\begin{eqnarray}
{\cal W}_i \prod_{j=1}^S \left( \frac{{\cal W}_j}{{\cal W}_j + 1} \right)^{2m}
= e^{x_i} \ .
\label{omegai}
\end{eqnarray}
The quantity $\Lambda$ is given by the series
\begin{eqnarray}
\Lambda = \lambda_0 + \lambda_1\sum_{j=1}^S {\cal W}_j +
\lambda_2 \sum_{j}\sum_{k<j} {\cal W}_j{\cal W}_k + \cdots
\label{Lambda}
\end{eqnarray}
whose coefficients are ($l=0,1,2,\cdots,S-1$)
\begin{eqnarray}
\lambda_l =  (2m+1)^{S-l} - [2(S-l)m+1]  \ .
\label{lambda}
\end{eqnarray}

According to the equations above 
$\lim_{{\cal W}_i\rightarrow \infty} \eta_i = 0$
leaving us with $S-1$ species in the
system. In this case, equations (\ref{etai}) and (\ref{omegai}) will
automatically converge to represent a system with the shortage of one species.
However, due to the energy structure of the CFs there is a hierarchy in the
values of ${\cal W}_i$. The ratio ${\cal W}_{i+1}/{\cal W}_i$ can be
obtained from (\ref{omegai})
\begin{eqnarray}
\frac{ {\cal W}_{i+1} }{ {\cal W}_i } = e^{x_{i+1}-x_i} =
e^{\beta[\varepsilon_{i+1}(k) - \varepsilon_i(k)]} \ ,
\label{fracw}
\end{eqnarray}
where we have assumed common temperature and chemical potential for all species.
Therefore, we see that if $\varepsilon_i(k)$ is the energy of the ith qLL
then ${\cal W}_i < {\cal W}_{i+1}$.

If this model is intended to reproduce the behavior of CFs some caution
is necessary because the gap energies depend on the total 
density $N(\vec{x})$
self-consistently
\begin{eqnarray}
E_{\delta}= \hbar \omega_{eff} =
\frac{\hbar |e|}{m^* c} B [1-2m\sum_j N_j(\vec{x})] \ .
\label{gap}
\end{eqnarray}

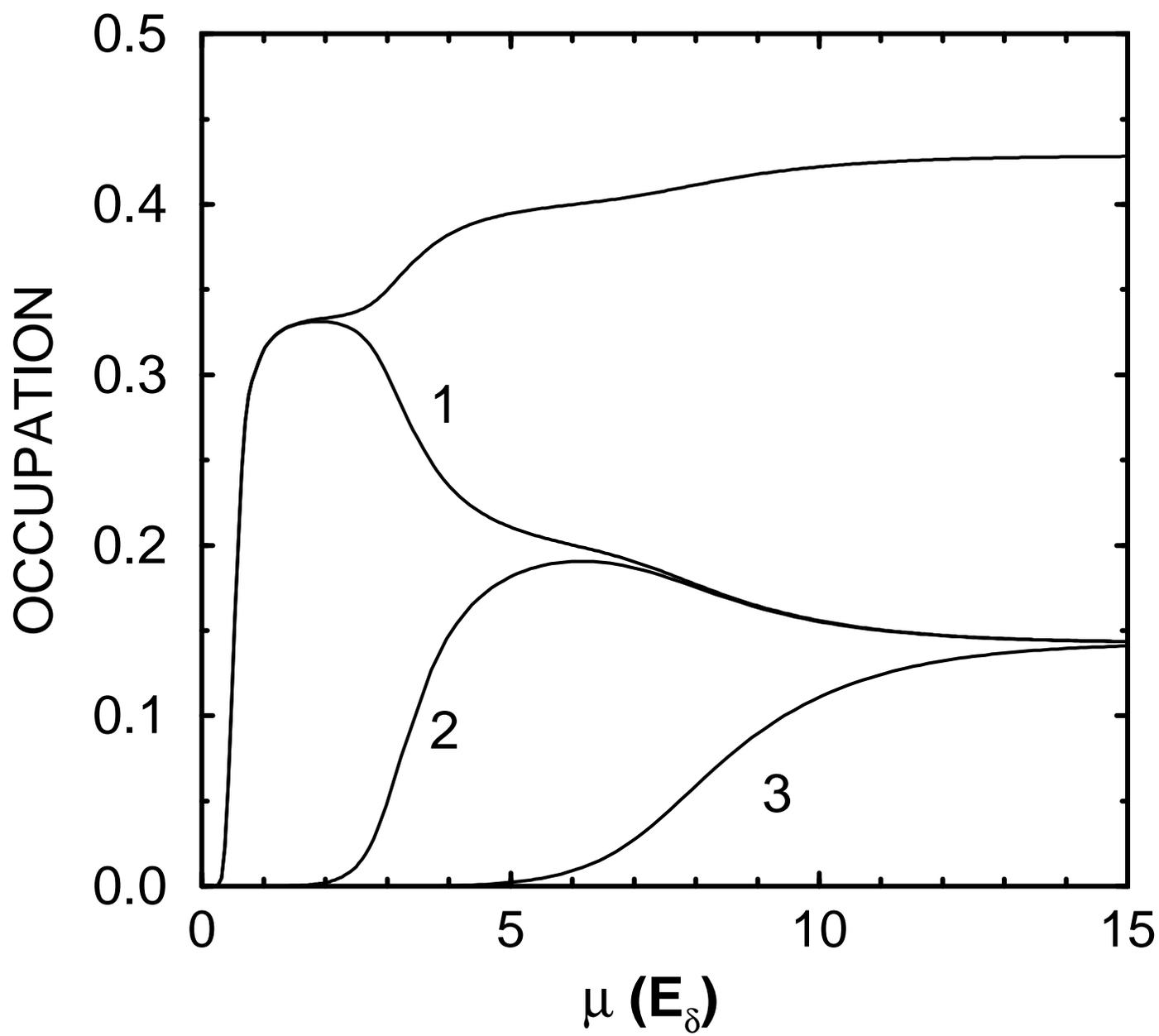
\begin{figure}
\caption{ Occupation values of the 3 lowest qLLs as a function of the 
generalized chemical potential. The indices indicate the occupation 
due to each level and the highest curve is the total occupation. 
The energy of the qLLs are $E_1=1/2 E_{\delta}$,
$E_2 = 3/2 E_{\delta}$ and $E_3 = 5/2 E_{\delta}$.  
The occupation and $E_{\delta}$ are obtained self-consistently for 
$m=1$. }
\label{occupation}
\end{figure}

In Figure \ref{occupation} we represent the occupation values of the 3 lowest
qLLs as a function of the generalized chemical potential $\mu$
of the FES particles.
The parameter $m=1$ and the density is assumed to be homogeneous. 
The occupation values were calculated self-consistently using
equations (\ref{etai}), (\ref{omegai}) and (\ref{gap}). The generalized chemical
potential is given in units of $E_{\delta}$ and the temperature is defined
by $\beta E^0_{\delta} = 30$, where $E^0_{\delta}$ is given by (\ref{gap})
when $N_j(\vec{x}) = 0$. We see that the occupation values have plateaus
that correspond to the fractions $1/(2pm+1)$.
These become poorly defined as the chemical potential  raises 
since $E_{\delta}$ decreases with the increasing density.

Having defined the statistical properties of this new particles we procced
with the calculation of the transport coefficients in the next section.

\subsection{Transport Coefficients}

The transport coefficients $L_{ij}$ for any filling factor 
$\nu = p/(2pm+1)$ can be obtained numerically using the 
distributions $\eta_i$, these being calculated self-consistently
with the gap energy. Nonetheless, for degenerate conditions analytical 
solutions are possible.

The result of section \ref{ss} for the Laughlin fractions 
can be obtained from the general formalism
above when we make ${\cal W}_i \rightarrow \infty$ for all $i \not = 1$.
In this limiting case 
\begin{eqnarray}
\eta_1 = \frac{1}{{\cal W}_1 + 2m + 1}
\label{tc1}
\end{eqnarray}
with $g=2m+1$. 

We now consider the situation in which two qLLs are populated. This means that 
$\eta_i=0$ for $i>2$  and 
\begin{eqnarray}
\eta_1 &=& \frac{{\cal W}_2 + 1}{({\cal W}_1+2m+1)({\cal W}_2+2m+1) - 4m^2}
\label{2s1} 
\\ \nonumber \\
\eta_2 &=& \frac{{\cal W}_1 + 1}{({\cal W}_1+2m+1)({\cal W}_2+2m+1) - 4m^2}
\label{2s2}
\end{eqnarray}
where $\eta_1$ describes the lowest band and $\eta_2$ the highest one.
Relation (\ref{fracw}) is now
\begin{eqnarray}
{\cal W}_1 = {\cal W}_2 e^{-\beta E_{\delta}(k)} \ ,
\label{2s41}
\end{eqnarray}
where $E_{\delta}(k)$ is the wave vector dependent spectral gap
between the bands
\begin{eqnarray}
E_{\delta}(k) = \epsilon_2(k) - \epsilon_1(k)  \ .
\label{2s42}
\end{eqnarray}
For positive $E_{\delta}(k)$, at low temperatures,
$\beta E_{\delta}(k) \gg 1$ that leads to the other important relation
${\cal W}_1 \ll {\cal W}_2$.

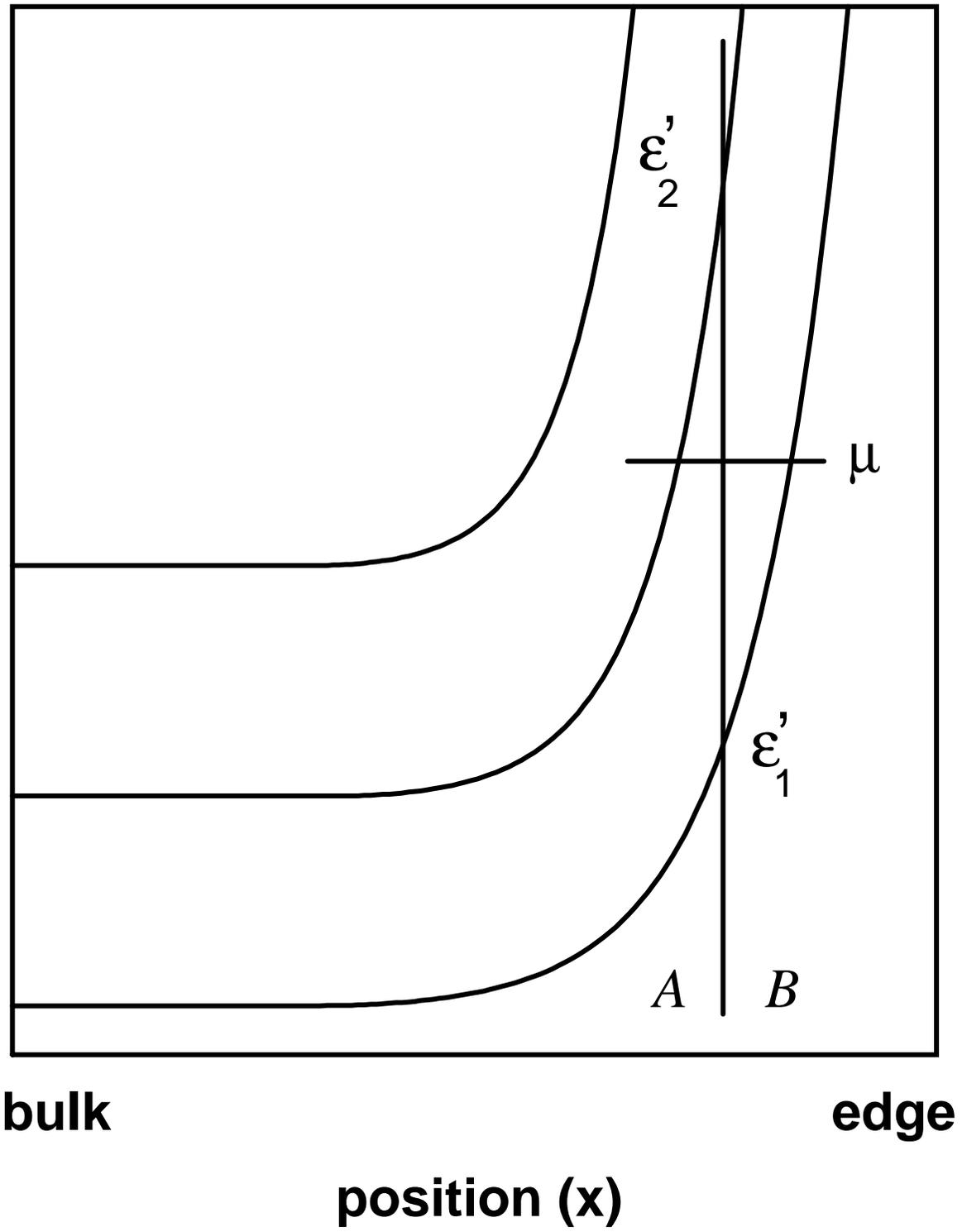
\begin{figure}
\caption{This scheme of the band structure shows the energy dispersion 
of the three lowest bands as a function of position, from the bulk
towards the right edge of the sample. The generalized chemical potential 
is indicated by the horizontal line. The vertical line is a reference.
The energy values $\epsilon'_2$ and $\epsilon'_1$ indicate the points 
where this reference line crosses the bands. }
\label{sample}
\end{figure}

In Figure \ref{sample}, a schematic band structure 
shows the energy dispersion of the three lowest bands as
a function of position, from the bulk
towards the right edge of the sample. 
The generalized chemical potential is indicated by
the horizontal line. The vertical line is a reference, it divides the band
structure into regions ${\cal A}$ and ${\cal B}$, whose meaning will be
discussed bellow. The energy values
$\epsilon'_2$ and $\epsilon'_1$ indicate the points where this reference line
crosses the bands.
We assume that the system is degenerate, so that
$\beta \epsilon'_1 \ll \beta \mu$ and $\beta \epsilon'_2 \gg \beta \mu$.
Notice that the generalized chemical potential ($\mu$) is above the 
lowest unoccupied state and lies over the third qLL band. This does not 
mean, however, that this level is populated. It occurs because the 
statistical mutual interaction modifies the electronic 
chemical potential.

Because of the mutual statistical interaction, the occupation functions
$\eta_i$ depend on both $\epsilon_2(k)$ and $\epsilon_1(k)$. We take advantage
of the degenerate condition of the system to circumvent this difficulty.
Since composite fermions at the same position in space have the same $k$,
if $k_B T \ll (\epsilon'_2 - \epsilon'_1)$ each ${\cal W}_i$ should go from 0 to
$\infty$ at different values of $k$, although around the same energy ($\mu$).
In other words, the transitions of ${\cal W}_i$ from 0 to $\infty$ are 
decoupled.
We have \\
{\it region} ${\cal A}$ : ${\cal W}_1 \ll 1$ while ${\cal W}_2$ 
makes the transition $0 \rightarrow \infty$, \\
{\it region} ${\cal B}$ : ${\cal W}_2 \gg 1$ while ${\cal W}_1$ 
makes the transition $0 \rightarrow \infty$.\\

Therefore the following approximations are possible. Focusing initially on
the highest band, in region ${\cal A}$
\begin{eqnarray}
\eta_2(k) = \frac{1}{(2m+1){\cal W}_2 + 4m +1}
\label{eta2A} \\
\nonumber \\
\frac{{\cal W}_2^{4m+1}}{({\cal W}_2+1)^{2m}} 
= e^{x_2(k) + 2m\beta E_{\delta}(k)} \ ,
\label{w2a}
\end{eqnarray}
while $\eta_2=0$ in region ${\cal B}$. 
So, integration over region ${\cal A}$ is enough to give us $I_2$ 
and $\dot{U}_2$ for the highest edge state.

For the lowest band the entire energy dispersion ought to be considered. In
region ${\cal A}$
\begin{eqnarray}
\eta_1^{{\cal A}}(k) = \frac{{\cal W}_2 + 1}{(2m+1){\cal W}_2 + 4m +1} 
\label{qwe} 
\end{eqnarray}
with ${\cal W}_2(k)$ given by (\ref{w2a}). In region ${\cal B}$
\begin{eqnarray}
\eta_1^{{\cal B}}(k) = \frac{1}{{\cal W}_1 + 2m +1} 
\label{poi} \\
\nonumber \\
\frac{{\cal W}_1^{2m+1}}{({\cal W}_1+1)^{2m}} = e^{x_1(k)} \ .
\label{lkj}
\end{eqnarray}
Therefore, for this band we have
$I_1 = I_1^{{\cal A}} + I_1^{{\cal B}}$ and
$\dot{U}_1 = \dot{U}_1^{{\cal A}} + \dot{U}_1^{{\cal B}}$.

Consider the current due to the highest band.
\begin{eqnarray}
I_2 = q \ \int_0^{\infty} \ \frac{dk}{2\pi} v_2(k) \eta_2(k) \ ,
\label{2s6}
\end{eqnarray}
with $\eta_2$ given by (\ref{eta2A}).
It is convenient to
define the new variable $E_2(k) = \epsilon_2(k) + 2m E_{\delta}(k)$,
in terms of which we write the velocity
\begin{eqnarray}
v_2(k) = \frac{1}{\hbar} \ \frac{d\epsilon_2}{dk} =
\frac{1}{\hbar} \left( \frac{dE_2}{dk} - 2m\frac{dE_{\delta}}{dk}
\right) \ .
\label{2s9}
\end{eqnarray}
Using $E_2$ as the variable of integration $I_2$  is rewritten as
\begin{eqnarray}
I_2 = \frac{q}{h}\ \int_{E_2^0}^{\infty} \left[ 1 -
2m\frac{dE_{\delta}}{dE_2} \right] \eta_2(E_2) dE_2 \ .
\label{2s10}
\end{eqnarray}
The lower limit of integration $E_2^0 = \epsilon_2^0 + 2mE_2^{bulk}$ is
a constant, with $E_2^{bulk}$ representing the value of the gap
in the bulk of the sample.

Another transformation eliminates the explicit dependence on the energy.
From Eq.(\ref{w2a}) we obtain $E_2$ as a function of ${\cal W}_2$ 
\begin{eqnarray}
\frac{dE_2}{d{\cal W}_2} = \frac{1}{\beta F}\ \frac{dF}{d{\cal W}_2} \ ,
\label{2s11}
\end{eqnarray}
with
\begin{eqnarray}
F({\cal W}_2) \equiv \frac{{\cal W}_2^{4m+1}}{({\cal W}_2+1)^{2m}} \ .
\label{2s12}
\end{eqnarray}

The gap $E_{\delta}$ can also be written in terms of ${\cal W}_2$. With this
purpose we return to the CF picture that gives us
$E_{\delta} = \hbar \omega_{eff} = \hbar (eB_{eff}/m^*c)$.
The effective
magnetic field depends on the local density of particles and so does the gap.
Therefore
\begin{eqnarray}
E_{\delta}(k) &=& \frac{\hbar |q|}{m^*c} B \left[ 1 - 2m\nu \right] \\
&=& E_{\delta}^0 \left[ 1 - 2m (\eta_1(k) + \eta_2(k)) \right] \ ,
\label{2s13}
\end{eqnarray}
where we have defined $E_{\delta}^0 = (\hbar |q|B/m^*c)$.
In the limit ${\cal W}_1 \ll 1$ which characterizes region ${\cal A}$
\begin{eqnarray}
\eta_1+\eta_2 \approx \frac{{\cal W}_2 +2}{(2m+1){\cal W}_2 + 4m+1} \ .
\label{2s14}
\end{eqnarray}

We are now able to write $I_2$ in a form that is independent of the details
of the particle spectra. Substituting  expressions (\ref{2s11}) to (\ref{2s14})
into Eq.(\ref{2s10}) we obtain
\begin{eqnarray}
I_2 = \frac{q}{h\beta} \lim_{{\cal W}_2^0 \rightarrow 0}  
\int_{{\cal W}_2^0}^{\infty}
& &\left[ \frac{1}{F}\frac{dF}{d{\cal W}_2} \right.
\nonumber \\   
& & + \left. a_m\frac{d(\eta_1+\eta_2)}{d{\cal W}_2} \right]
\eta_2({\cal {\cal W}}_2) d{\cal W}_2 \ ,
\label{2s15}
\end{eqnarray}
where $a_m = 4m^2\beta E^0_{\delta}$.
Separating this expression, the integration of the first part gives us
\begin{eqnarray}
\frac{q}{h\beta} \ \int_{{\cal W}_2^0}^{\infty}
\ \frac{\eta_2}{F}\frac{dF}{d{\cal W}_2}
d{\cal W}_2 & &= \nonumber \\
& &\frac{q}{h\beta} \left[\ln{({\cal W}_2^0+1)} - 
\ln{({\cal W}_2^0)} \right] \ , 
\label{2s16}
\end{eqnarray}
whereas the integration of the second term produces a
quantity independent of temperature and chemical potential when we make
${\cal W}_2^0=0$,
that will have no contribution to the transport coefficients.
The limit ${\cal W}_2^0 \rightarrow 0$ of expression (\ref{2s16}) 
can be obtained
from (\ref{w2a}) that gives us
\begin{eqnarray}
{\cal W}_2^0 = e^{\frac{\beta(E_2^0 -\mu)}{4m+1}} 
({\cal W}_2^0 +1)^{\frac{2m}{4m+1}} \ .
\label{2s17}
\end{eqnarray}
Substituting this result in Eq.(\ref{2s16}) the limit 
${\cal W}_2^0 \rightarrow 0$ of $I_2$
can be easily obtained. It depends only on the chemical potential and,
therefore, the coefficients that describe the electrical current for 
the highest band are
\begin{eqnarray}
\frac{\partial I_2}{\partial \mu} = \frac{1}{4m+1}\frac{q}{h}  \ \ \ \ \ \ \ ,
\ \ \ \ \ \ \ \frac{\partial I_2}{\partial T} = 0 \ .
\label{2s18}
\end{eqnarray}

The electrical current $I_1$ due to the lowest band is obtained by 
same approach.
We perform the integration
dividing the whole domain in two regions (see Fig.\ref{sample}).
Along the first portion, designated by ${\cal A}$, 
${\cal W}_1 \ll 1$ throughout the
range whereas ${\cal W}_2$ increases from ${\cal W}_2 \ll 1$ to 
${\cal W}_2 \gg 1$.
In the second part, represented by ${\cal B}$, ${\cal W}_2$ 
remains very large while
${\cal W}_1$ makes the transition ${\cal W}_1 \ll 1$ to ${\cal W}_1 \gg 1$.

Along region ${\cal A}$ the occupation  
$\eta_1$ is given by (\ref{qwe}) and 
group velocity for this band is
\begin{eqnarray}
v_1(k) = \frac{1}{\hbar} \ \frac{d\epsilon_1}{dk} =
\frac{1}{\hbar} \left( \frac{dE_2}{dk} - (2m+1)\frac{dE_{\delta}}{dk}
\right) \ .
\label{2s19}
\end{eqnarray}
Using the Eq.(\ref{2s19}) along with the identities
(\ref{2s11}) to (\ref{2s14}) we are able to write
\begin{eqnarray}
I_1^{\cal A} &=& \frac{q}{h}\ \int_{\cal A}
\left[ 1 - (2m+1)\frac{dE_{\delta}}{dE_2} \right] \eta_1^{\cal A}(E_2) dE_2
\\
&=& \frac{q}{h\beta} \lim_{{\cal W}_2^0 \rightarrow 0}  
\int_{{\cal W}_2^0}^{{\cal W}'_2}
\left[ \frac{1}{F}\frac{dF}{d{\cal W}_2}  \right.
\nonumber \\    
& & \ \ \ \ \ \ \ \ \ \ \ \ \ \ \
+ \left. b_m\frac{d(\eta_1+\eta_2)}{d{\cal W}_2} \right]
\eta_1^{{\cal A}}({\cal W}_2) d{\cal W}_2 \ ,
\label{2s21}
\end{eqnarray}
where $b_m = 2m(2m+1)\beta E^0_{\delta}$ and 
${\cal W}'_2 = {\cal W}(\varepsilon'_2)$ 
is independent of the thermodynamical parameters for degenerate systems.

Once again the second term in the integrand will produce a quantity that is
independent of temperature and chemical potential, therefore irrelevant for
the transport coefficients. On the other hand, the first term is 
\begin{eqnarray}
\frac{q}{h\beta} \ \int_{{\cal W}_2^0}^{{\cal W}'_2}
\ \frac{\eta^{{\cal A}}_1}{F}\frac{dF}{d{\cal W}_2}
d{\cal W}_2 = \frac{q}{h\beta} \left[\ln{({\cal W}'_2)} - 
\ln{({\cal W}_2^0)} \right] \ .
\end{eqnarray}
Then using (\ref{2s17}) we get
\begin{eqnarray}
\frac{\partial I_1^{\cal A}}{\partial \mu} = \frac{1}{4m+1}\frac{q}{h}   
\ \ \ \ \ \ \ , \ \ \ \ \ \ \
\frac{\partial I_1^{\cal A}}{\partial T} = 0 \ .
\label{2s22}
\end{eqnarray}

As for region ${\cal B}$, its contribution to the current due to the lowest
band is 
\begin{eqnarray}
I_1^{{\cal B}} = \frac{q}{h\beta}\ \int_{{\cal W}'_1}^{\infty}
\frac{\eta_1^{{\cal B}}}{H}\frac{dH}{d{\cal W}_1} d{\cal W}_1 \ ,
\label{2s24}
\end{eqnarray}
with 
\begin{eqnarray}
H({\cal W}_1) = \frac{ {\cal W}_1^{2m+1} }{ ({\cal W}_1+1)^{2m} } \ .
\end{eqnarray}
The lower limit ${\cal W}'_1 = {\cal W}_1(\varepsilon'_1)$
corresponds to a continuation from ${\cal W}'_2$ (see Fig. \ref{sample}).
The differentiation of (\ref{2s24}) with respect to the thermodynamical 
parameters vanishes.

Therefore we sum up the results of this calculation as 
\begin{eqnarray}
L_{11} &=& \frac{\partial (I_1 + I_2)}{\partial \mu} = \frac{2}{4m+1}\frac{q}{h}
\\
L_{12} &=& 0 \ .
\end{eqnarray}

The calculations leading to the coefficients $L_{21}$ and $L_{22}$ are not 
presented since they follow the same formalism. However we 
write the final results 
\begin{eqnarray}
\delta I &=& \frac{2}{4m+1} \frac{q}{h} \delta \mu
\label{e29}
\\
\delta \dot{U} &=& \left\{ \frac{2}{4m+1} \frac{\mu}{h}  +
C_m \frac{E_{\delta}}{h} \right\} \delta \mu
+ 2\ \frac{\pi^2}{3}\frac{k_B^2 T}{h} \delta T \ ,
\label{e30}
\end{eqnarray}
where $C_m$ is a non-universal coefficient that depends on $m$.
Although $\mu$ differs from the electrochemical potential 
$\delta \mu$ does not when the system is in a FQHE plateau,
as evidenced by (\ref{e29}).
Despite the statistical coupling between the two CF quasi-Landau levels,
once again the universal quantum of thermal conductance is obtained, this
time multiplied by 2, which reflects the presence of the two
modes of propagation.
The two-terminal electrical conductance
$G = 2/(4m+1)(e^2/h)$ is also obtained for this family of states, in agreement
with experiment.

\section{Conclusions}

In conclusion, we have presented a generalized theory
of transport of 1D systems.
We have shown that the ballistic thermal conductance of
one-dimensional systems is statistics-independent and
thus truly universal:
$\kappa^{univ} = \frac{\pi^2}{3}(k_B^2 T/h)$. This result is valid in the
degenerate regime for systems of particles obeying
fractional exclusion statistics, whether they present a Fermi surface
or are described by a Planck distribution.
Electrical conductances for ballistic electrons in 1D quantum wires
and in the FQH regime, although not universal (in the sense that they
depend on statistics), also follow naturally from this theory.

\section{Acknowledgements}

This work was supported by NSERC.

\end{multicols}


\end{document}